\documentclass[floats,prd,aps,showpacs,amssymb,nofootinbib]{revtex4}
\usepackage{amssymb}
\usepackage{graphics}
\usepackage{amsmath}
\usepackage{amsfonts}
\usepackage{bm}
\usepackage[]{latexsym}

\newcommand{\be}{\begin{equation}}\newcommand{\ee}{\end{equation}}
\newcommand{\bea}{\begin{eqnarray}}\newcommand{\eea}{\end{eqnarray}}
\newcommand{\brr}{\begin{array}}\newcommand{\err}{\end{array}}
\newcommand{\bit}{\begin{itemize}}\newcommand{\eit}{\end{itemize}}
\newcommand{\ben}{\begin{enumerate}}\newcommand{\een}{\end{enumerate}}

\newcommand{\ba}{\begin{array}}
\newcommand{\ea}{\end{array}}

\def\al{\alpha}

\def\1{{_{1}}}\def\2{{_{2}}}

\def\noHe0{:\;\!\!\;\!\!:H_e(0):\;\!\!\;\!\!:}
\def\noHm0{:\;\!\!\;\!\!:H_\mu(0):\;\!\!\;\!\!:}
\def\al{\alpha}

\def\1{{_{1}}}\def\2{{_{2}}}

\begin{document}
\title{Neutrino oscillations in accelerated frames}
\author{M Blasone\footnote{blasone@sa.infn.it}$^{\hspace{0.3mm}1,2}$, G Lambiase\footnote{lambiase@sa.infn.it}$^{\hspace{0.3mm}1,2}$, G G Luciano\footnote{gluciano@sa.infn.it}$^{\hspace{0.3mm}1,2}$ and L Petruzziello\footnote{lpetruzziello@na.infn.it}$^{\hspace{0.3mm}1,2}$} \affiliation
{\vspace{1mm}$^1$Dipartimento di Fisica, Universit\'a di Salerno, Via Giovanni Paolo II, 132 I-84084 Fisciano (SA), Italy.
\\ 
\vspace{1mm}
$^2$INFN, Sezione di Napoli, Gruppo collegato di Salerno, Italy.}

\date{\today}
  \def\be{\begin{equation}}
\def\ee{\end{equation}}
\def\al{\alpha}
\def\bea{\begin{eqnarray}}
\def\eea{\end{eqnarray}}

\begin{abstract}

We discuss neutrino oscillations in vacuum from the point of view of a uniformly accelerated
observer. A covariant definition of quantum phase is introduced with the aim of generalizing the standard expression of the oscillation amplitude to the accelerating frame. By way of illustration, we address a simplified two-flavor model with relativistic neutrinos, showing that inertial effects on the usual Pontecorvo formula are intimately related to the energy redshift. Phenomenological aspects are preliminarily analyzed in the context of atmospheric neutrinos. Finally, we discuss a gedanken experiment in order to investigate our formalism in regime of extreme acceleration.
\end{abstract}

 \vskip -1.0 truecm
\maketitle

\section{Introduction}
\setcounter{equation}{0}
Neutrino oscillations in  flat spacetime have been extensively analyzed
since Pontecorvo's pioneering idea of non-degenerate mass-matrix~\cite{Pontecorvo}.
Over the years, however, alternative mechanisms have
been proposed: among these, worthy of note are
the ones suggested by Gasperini~\cite{Gasperini:1989rt}
and Liu~\cite{Liu:1997km}, respectively.
Although they have  both been rejected by experiments, these solutions
represent a first attempt to accommodate gravitational
effects into the standard picture of neutrino oscillations.
A systematic treatment of flavor oscillations in curved spacetime has
been discussed by a number of authors in
Refs.~\cite{Aluw,Ahluwalia:1996ev,Cardall}.
In Ref.~\cite{Cardall}, in particular,
the authors introduce a simple formalism to demonstrate
that gravitational effects are closely related to the redshift of
neutrino energy.
The framework becomes even richer in astrophysical regimes, where the presence of strong gravitational and magnetic fields (provided that neutrinos possess a non-vanishing magnetic moment) may significantly affect the oscillation probability~\cite{Aluw,Lambiase:2004qk}.
Due to the equivalence principle, similar results are expected to be valid also in accelerated frames.
Along this line, a pilot analysis
of phenomenological aspects of neutrino oscillations for an accelerating and rotating observer
has been performed in Ref.~\cite{Capozziello:1999ww}.
Recently, mixing transformations in Rindler (uniformly accelerated) background
have been also studied in Quantum Field Theory (QFT)~\cite{Blasone:2017nbf,Blasone:2018byx}, showing that non-thermal corrections to the Unruh radiation may arise due to the interplay between the Bogoliubov transformation related to the structure of Rindler spacetime and the one hiding in field mixing~\cite{Blasone:1995zc}.

Apart from phenomenological implications, we stress that a deeper
understanding of inertial effects on flavor mixing and oscillations
may shed some light on a number of intriguing issues
at a theoretical level. Recently, indeed, the role of neutrino mixing
in the decay of accelerated protons (inverse $\beta$-decay) has been investigated
with controversial results~\cite{Ahluwalia:2016wmf,Blasone:2018czm,Cozzella:2018qew}.
Specifically, in Refs.~\cite{Matsas:1999jx} it was pointed out that the Unruh effect is necessary  to maintain the general covariance of QFT when considering the
inverse $\beta$-decay rate in the laboratory and comoving
frames, respectively. Subsequently, it was noted that
neutrino mixing can spoil this agreement~\cite{Ahluwalia:2016wmf}, and
further discussion~\cite{Blasone:2018czm,Cozzella:2018qew}
has narrowed down possible causes to the effective nature of asymptotic neutrino states
as mass or flavor eigenstates. Cleary, such an
ambiguity affects flavor oscillations too.
In particular, since the oscillation probability calculated in the ordinary QFT
by means of the exact flavor states~\cite{Blasone:1998hf} contains extra-terms
with respect to the usual quantum mechanical formula,
one expects corrections to arise also for the non-inertial case.

In the present work, a preliminary step
along this direction is taken by analyzing the effects of a linear acceleration on the
neutrino oscillation
formula in the context of Quantum Mechanics (QM).
The obtained result should thus be regarded as a benchmark
for the field theoretical treatment of the problem, for which work is in progress.

The paper is structured as follows: in Section II, we briefly review the standard treatment of neutrino oscillations in  flat spacetime using the plane-wave formalism.
Section III is devoted to a heuristic derivation of the oscillation probability  for a uniformly accelerated observer. The same result is recovered in Section IV
by solving the Dirac equation in accelerated  frames. The obtained expression is critically compared with the one in Ref.~\cite{Capozziello:1999ww}, where corrections are calculated in a more geometric framework.
As possible applications, in Section V we discuss how Earth's gravity affects the oscillation probability of atmospheric neutrinos. In addition, we propose a gedanken experiment in which an ideal detector is used for testing inertial effects in proximity of high-density astrophysical objects. Section VI contains conclusions and an outlook at future developments of the present work.

Throughout the paper, we shall use natural units $\hslash=c=1$  and the flat Minkowski metric with the conventional timelike signature
\begin{equation}
\eta_{\mu\nu}\, =\, {\rm diag}(+1,-1,-1,-1)\,.
\label{eqn:flatmetr}
\end{equation}

\section{Neutrino oscillations in  flat spacetime}
We start by reviewing the standard theory
of neutrino oscillations in Minkowski spacetime.
For the sake of simplicity, we focus on a
 model with only two flavor generations (for a more rigorous three-flavor description, we remand the reader to Ref.~\cite{Maki:1962mu}).

In the conventional matrix notation, indicating by $|\nu_\alpha\rangle$ ($\alpha=e,\mu$) and
 $|\nu_k\rangle$ ($k=1,2$) neutrino flavor and mass
 eigenstates, respectively, the following relation holds~\cite{KimPev}
\begin{equation}
\label{eqn:U}
\begin{pmatrix}
|\nu_e\rangle  \\
|\nu_\mu\rangle
\end{pmatrix}\,=\,U(\theta)\begin{pmatrix}
|\nu_1\rangle  \\
|\nu_2\rangle
\end{pmatrix},
\end{equation}
where $\theta$ is the mixing angle and $U(\theta)$ is the Pontecorvo  unitary
matrix
\begin{equation}
\label{PMM}
U(\theta)\,=\,
\begin{pmatrix}
\cos\theta&\sin\theta\\
-\sin\theta&\cos\theta
\end{pmatrix}\,.
\end{equation}
In what follows, we describe the propagation of the
mass eigenstates by plane-waves, i.e.
\begin{equation}
|\nu_k(t,\textbf{x})\rangle\,=\,\exp[-i\Phi_k(t, \textbf{x})]\,|\nu_k\rangle,\quad (k\,=\,1,2)\,,
\end{equation}
where
\begin{equation}
\Phi_k\,=\,E_k\,t-\textbf{p}_k\cdot\textbf{x}
\end{equation}
is the quantum-mechanical phase
of the $k^{th}$ neutrino state, with $E_k$ and $\textbf{p}_k$ being its energy and momentum, respectively. Mass, energy and momentum are related by the mass-shell condition
\begin{equation}
E^2_k\,=\,m^2_k\,+\,{|\textbf{p}_k|}^2\,.
\end{equation}
In the relativistic approximation, labelling with $A\hspace{0.1mm}(t_A, \textbf{x}_A)$ and $B\hspace{0.1mm}(t_B, \textbf{x}_B)$  the spacetime points in which neutrinos are produced and
detected, respectively, the phase acquired by the $k^{th}$ eigenstate after propagating
over the distance  $L_p\,\equiv\,\left|\textbf{x}_B-\textbf{x}_A\right|$ reads\footnote{In order for the interference pattern not
to be destroyed, we remark that neutrinos
must be produced coherently and  measured at
the same spacetime point.}
\begin{equation}
\label{eqn:phase}
\Phi_{k}\,=\,E_k\hspace{0.1mm}(t_B-t_A)\,-\,|\textbf{p}_k| |\textbf{x}_B-\textbf{x}_A|\,\approx\,\frac{m^2_k}{2E_0}L_p\,.
\end{equation}
Notice that, in the second step of Eq.~(\ref{eqn:phase}), we have exploited the relativistic condition $m_k\ll  E_k$, so that
\begin{equation}
\label{eqn:approx}
t_B\,-\,t_A\,\simeq\,\left|\textbf{x}_B\,-\,\textbf{x}_A\right|,
\end{equation}
and the first order expansion for $E_k$
\begin{equation}
\label{ap}
E_k\,\simeq\, E_0 \,+\, \mathcal {O}\left(\frac{m^2_k}{2E_0}\right)\,,
\end{equation}
with $E_0$ being the energy for a massless
neutrino. The last equation amounts to require that mass eigenstates are also energy eigenstates with a common energy $E_0$.

Let us now consider an electron neutrino $|\nu_e\rangle$ emitted via weak interaction at the  point $A\hspace{0.1mm}(t_A, \textbf{x}_A)$.  Using Eq.~(\ref{eqn:U}), the probability that it is revealed as muon neutrino $|\nu_\mu\rangle$ at the point $B\hspace{0.1mm}(t_B, \textbf{x}_B)$ is given by
\begin{eqnarray}
\label{pontec}
\nonumber
P_{\nu_e\rightarrow\nu_\mu}\,&\equiv&\,{\big|\langle\nu_\mu(t_B, \textbf{x}_B)\hspace{0.1mm}|\hspace{0.1mm}\nu_e(t_A, \textbf{x}_A)\rangle\big|}^2\\[2mm]
\,&=&\,\sin^2(2\theta)\sin^2\left(\frac{\Phi_{12}}{2}\right),
\end{eqnarray}
where, according to Eq.~(\ref{eqn:phase}), the phase-shift $\Phi_{12}\equiv \Phi_1-\Phi_2$ takes the form
\begin{equation}
\label{eqn:relativephase}
\Phi_{12}\,\simeq\, \frac{\Delta m^2_{12}}{2E_0}\hspace{0.2mm}L_p\hspace{0.1mm}\,\equiv\,  \frac{m_1^2\,-\,m_2^2}{2E_0}\hspace{0.2mm}L_p\hspace{0.1mm}\,.
\end{equation}
It should be noted that, in the case where at least one
of the states $|\nu_k\rangle$ is non-relativistic, a wave packet approach is required instead of the above plane-wave formalism~\cite{Giunti:1991ca}.
For our purposes, however, such an analysis would show that the
approximations Eqs.~(\ref{eqn:approx}) and (\ref{ap}) are adequate, leading to the
formula Eq.~\eqref{pontec} for the oscillation probability. We also stress that this equation
represents the quantum-mechanical limit of a more general formula derived within the QFT framework. For a detailed analysis of this, see Refs.~\cite{Blasone:1995zc}.

The foregoing discussion applies to an observer at rest or moving inertially
with respect to the oscillation experiment. Nevertheless, due to gravity,
any stationary laboratory on Earth experiences a
linear acceleration (in the present analysis, we do not take care of rotational effects. A thorough discussion of this subject can be found in
Ref.~\cite{Capozziello:1999ww}). To show how acceleration
affects flavor oscillations,
let us then recast the quantum mechanical phase Eq.~\eqref{eqn:phase}
into a covariant form, according to~\cite{Stodolsky:1978ks}
\begin{equation}
\Phi_k\,=\,\int_A^{B}p_\mu^{(k)}\hspace{0.1mm}dx^{\mu},
\label{eqn:covphase}
\end{equation}
where
\begin{equation}
\label{eqn:moment}
p_\mu^{(k)}\,=\,m_k\hspace{0.3mm}g_{\mu\nu}\hspace{0.3mm}\frac{dx^\nu}{ds}
\end{equation}
is the canonical four-momentum conjugated to the coordinates $x^\mu$
and $ds$, $g_{\mu\nu}$ are the line element and the metric
tensor, respectively. The integration in Eq.~\eqref{eqn:covphase}
has to be performed  along the light-ray trajectory linking the spacetime points
$A$ and $B$. For $g_{\mu\nu}$ corresponding to the flat
metric Eq.~\eqref{eqn:flatmetr}, it is easy to show that Eqs.~(\ref{eqn:covphase}) and (\ref{eqn:moment})
 reproduce the standard result Eq.~\eqref{eqn:phase}, as it should be.

\section{Inertial effects on neutrino oscillations: a heuristic treatment}

We now turn to the discussion of neutrino oscillations for a uniformly accelerated
observer. In order to apply the covariant formalism above described, let us recall that the line element in an accelerated frame can be written as (we neglect the effects of the spacetime curvature)~\cite{Misner}
\begin{equation}
\label{eqn:linelementacframe}
ds^2\,=\,f(\textbf{a},\textbf{x})\hspace{0.1mm}\hspace{0.1mm}{(dt)}^2
\,-\,d\textbf{x}\cdot d\textbf{x}\,,
\end{equation}
where
\begin{equation}
\label{eqn:funct}
f(\textbf{a},\textbf{x})\,\equiv\,\left(1\,+\,\textbf{a}\cdot\textbf{x}\right)^2\,,
\end{equation}
with $\textbf{a}$ being the proper three-acceleration and  $x^\mu=(t,\textbf{x})$
the \emph{Fermi coordinates} for an accelerated observer~\cite{Misner,lrf}, whose range of validity is limited by the requirement $|\textbf{x}|\ll {|\textbf{a}|}^{-1}$. This occurs because the above reference frame is conceived to describe a neighborhood of the observer's world line as long as the previous condition holds. However, the confinement on the spatial region does not affect the relevance of our considerations, since typical oscillation lengths of neutrino experiments allow us to deal with even considerable values of $\textbf{a}$. For instance, for acceleration of the order of Earth's gravity, the metric is valid within a range of one light-year.

\begin{figure*}[t]
\resizebox{18cm}{!}{\includegraphics{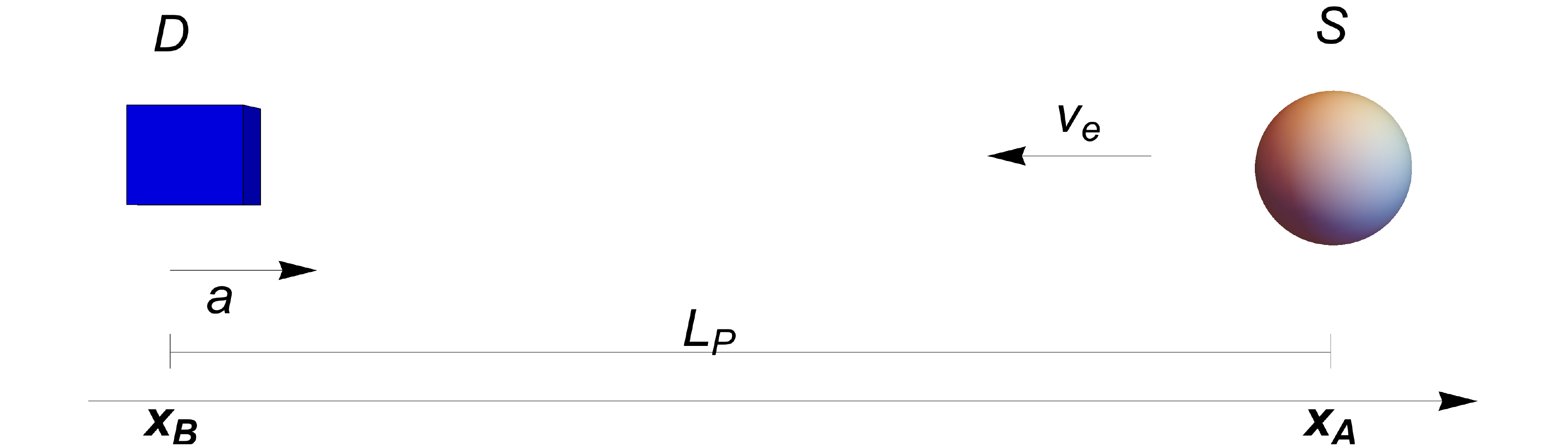}}
\caption{Neutrino emission from a source $S$ at the point $x_A$. After propagating
over the distance $L_p$, neutrinos reach the detector $D$ at the point $x_B\,<\,x_A$. Note that the direction of motion of neutrinos is assumed to be antiparallel to the acceleration of the detector.}
\label{fig1}
\end{figure*}

Without loss of generality, we can restrict our analysis to $1+1$ dimensions,
assuming the acceleration to be antiparallel to the direction of neutrino propagation (see Fig~\ref{fig1}). 
According to Eq.~\eqref{eqn:moment}, the components of the neutrino
canonical momentum $p_\mu^{(k)}$ are
\begin{eqnarray}
p_t^{(k)}&=&m_k\hspace{0.3mm}f(a, x)\hspace{0.2mm}\frac{dt}{ds}\,,
\\[1mm]
p_x^{(k)}&=&-m_k\frac{dx}{ds}\,.
\end{eqnarray}
They are related to each other and to the mass $m_k$ by the
generalized mass-shell condition
\begin{equation}
m^2_k\,=\,g^{\mu\nu}\hspace{0.1mm}p^{(k)}_\mu\hspace{0.1mm}p^{(k)}_\nu,
\label{eqn:genmasscond}
\end{equation}
with $g_{\mu\nu}$ given in Eq.~(\ref{eqn:linelementacframe}). Since the metric does not depend on the coordinate $t$, the timelike momentum component
$p_t^{(k)}$ is conserved along the geodesic trajectory
of the $k^{th}$ neutrino eigenstate.
We define such a constant to be $p_t^{(k)}\equiv E_k$.
It represents the energy measured by an observer at rest
at the origin. Due to the metric Eq.~(\ref{eqn:linelementacframe}), however,
it differs from the energy at any other
spacetime point. The local energy, defined as the energy measured by an
observer at rest at the generic position $x$, is related to $E_k$ by~\cite{Misner}
\begin{equation}
\label{eqn:localenergy}
E_k^{(loc)}(x)\,=\,\left|g_{tt}\right|^{-1/2}\hspace{0.1mm}E_k\,=\,{f(a, x)}^{-1/2}E_k\,.
\end{equation}
Next, by considering relativistic neutrinos and using Eq.~(\ref{eqn:covphase}), the phase of the $k^{th}$ neutrino eigenstate reads
\begin{equation}
\label{eqn:covariantphase}
\Phi_k\,=\,\int_{A}^{B}\left[E_k\left(\frac{dt}{dx}\right)_{\!0}\hspace{-0.5mm}\,-\,p_{k}(x)\right]dx\,,
\end{equation}
where the momentum  $p_k(x)\equiv- p_x{(k)}$ is obtained from the generalized mass-shell condition Eq.~(\ref{eqn:genmasscond}) as
\begin{equation}
\label{eqn:momentum}
p_k(x)\,=\,-{\left(\frac{E^2_k}{f(a, x)}\,-\,m^2_k\right)}^{1/2},
\end{equation}
and the light-ray differential $\left(dt/dx\right)_0$ is given by
\begin{equation}
\label{eqn:lightray}
\left(\frac{dt}{dx}\right)_{\!0}\,=\, -f(a, x)^{-1/2}\,,
\end{equation}
where the minus sign in Eqs.~(\ref{eqn:momentum}) and (\ref{eqn:lightray})
is due to the fact that neutrinos propagation is antiparallel to the $x$-axis.
By inserting Eqs.~(\ref{eqn:momentum}) and (\ref{eqn:lightray})
into Eq.~(\ref{eqn:covariantphase}), we get
\begin{equation}
\widetilde\Phi_k\,=\,-\int_{A}^{B}E_k\left\{1\,-\,{\sqrt{1-f(a, x)\frac{m^2_k}{E_k^2}}}\right\}{f(a, x)}^{-1/2}\hspace{0.3mm}dx\,,
\label{eqn:simplcovphase}
\end{equation}
where the tilde has been introduced to distinguish the above expression of the phase from the standard one in Eq.~(\ref{eqn:phase}).

Now, since detecting non-relativistic neutrinos is an extremely hard task,
it is reasonable to require that
\begin{equation}
\label{eqn:relativcond}
\frac{m^2_k}{{\left[E_k^{(loc)}(x_B)\right]}^2}\,=\,f(a, x_B)\frac{m^2_k}{E_k^2}\,\ll\, 1\,.
\end{equation}
This amounts to restrict our analysis to neutrinos that are relativistic
at the detector position $B\hspace{0.1mm}(t_B, x_B)$, and thus along all their path
$(x_B\le x\le x_A)$, for we have
\begin{eqnarray}
\label{eqn:localenergyrelation}
f(a, x)\frac{m^2_k}{E_k^2}\,&=&\,\frac{f(a, x)}{f(a, x_B)}
\frac{m^2_k}{{\big[E_k^{(loc)}(x_B)\big]}^2}\\[2mm]
\nonumber
&\,\equiv\,&\frac{1-ax}{1-ax_B}\,\frac{m_k^2}{{\big[E_k^{(loc)}(x_B)\big]}^2}\,\le\,\frac{m^2_k}{{\big[E_k^{(loc)}(x_B)\big]}^2}\,\ll\, 1\,.
\end{eqnarray}
Equations~(\ref{eqn:relativcond}) and (\ref{eqn:localenergyrelation}) allow
us to approximate the covariant phase Eq.~(\ref{eqn:simplcovphase})
as follows
\begin{equation}
\label{eqn:covphaseapprox}
\widetilde\Phi_k\,\simeq\,-\int_{A}^{B}\frac{m^2_k}{2\hspace{0.3mm}E_0}\hspace{0.2mm}{f(a, x)}^{1/2}dx\,,
\end{equation}
where, as in the absence of acceleration, we have used the first-order
approximation $E_k\,\simeq\, E_0\, +\,
\mathcal {O}\left(\frac{m^2_k}{2E_0}\right)$, with $E_0$ being the energy
at the origin for a massless particle.
Since this energy is constant along the light-ray trajectory between $A$ and $B$, the integration in Eq.~(\ref{eqn:covphaseapprox})
can be readily  performed, obtaining
\begin{equation}
\label{eqn:phaseapprox}
\widetilde\Phi_k\,\simeq\,\frac{m^2_k}{2E_0}|x_B\,-\,x_A|\left(1\,-\,\phi_a\right),
\end{equation}
where we have introduced the short-hand notation
\begin{equation}
\label{eqn:phia}
\phi_a\,\equiv\,\frac{a}{2}\left(x_B\,+\,x_A\right).
\end{equation}
Thus, the phase-shift responsible for the oscillation takes the form
\begin{equation}
\label{eqn:phaseshiftapprox}
\widetilde\Phi_{12}\,\simeq\,\frac{\Delta m^2_{12}}{2E_0}|x_B\,-\,x_A|\left(1\,-\,\phi_a\right)\,=\,\frac{\Delta m^2_{12}}{2E_0}\hspace{0.3mm}L_p\left(1\,-\,\phi_a\right),
\end{equation}
where we have used the definition of proper distance at constant time $dl\,\equiv\,\sqrt{-g_{xx}}\hspace{0.5mm}dx\,=\,dx$.

We remark that Eq.~(\ref{eqn:covphaseapprox})
does not match with the corresponding result Eq.~(25) of Ref.~\cite{Capozziello:1999ww}.
In that case, indeed, the correction to the neutrino phase-shift depends logarithmically on the acceleration.
We suspect that such a discrepancy arises because of an incorrect derivation of the final expression of the phase-shift in Ref.~\cite{Capozziello:1999ww} from the corresponding
formula in Ref.~\cite{Cardall}.

Now, in order to compare Eq.~(\ref{eqn:phaseshiftapprox})
with the standard result Eq.~(\ref{eqn:relativephase}),
let us rewrite $E_0$ in
terms of the neutrino local energy at the detector position $B\hspace{0.1mm}(t_B, \textbf{x}_B)$. Using Eq.~(\ref{eqn:localenergy}), it  follows that
\begin{equation}
\label{eqn:phaseshiftapprox2}
\widetilde\Phi_{12}\,\simeq\,\frac{\Delta m^2_{12}\,L_p}{2E^{(loc)}_0(x_B)\hspace{0.2mm}{f(a, x_B)}^{1/2}}\left(1\,-\,\phi_a\right).
\end{equation}
By virtue of the condition on the range of validity of the adopted metric (namely $a\,x\ll 1$), Eq.~(\ref{eqn:phaseshiftapprox2})
can be further manipulated, thus giving
\begin{equation}
\label{eqn:finalform}
\widetilde\Phi_{12}\,\simeq\,\frac{\Delta m^2_{12}\hspace{0.3mm}L_p}{2E^{(loc)}_0(x_B)}\left[1\,-\,\frac{a}{2}\hspace{0.2mm}L_p\right],
\end{equation}
where we have neglected higher order terms in the acceleration.
The first term on the r.h.s. is the
only surviving contribution for vanishing acceleration.
As expected, it corresponds
to the standard oscillation phase in Eq.~(\ref{eqn:relativephase}).
The remaining term provides the correction induced by a uniform, linear
acceleration on the neutrino oscillation phase.

\section{Inertial effects on neutrino oscillations: a geometric treatment}
In the previous section, we have derived inertial effects
on neutrino oscillations in a simple heuristic way. Using a more geometric treatment, we now want to prove that the same result  can be obtained by solving the Dirac equation
in an accelerated frame.  As a first step, let us write down the
covariant Dirac equation in curved spacetime~\cite{Weinberg}
\begin{equation}
\label{eqn:covdireq}
\Big[i\gamma^{a}\hspace{0.2mm}e^{\mu}_{a}\left(\partial_\mu\,+\,\Gamma_\mu\right)\,-\,M\Big]\psi\,=\,0\,,
\end{equation}
where $M$ is the neutrino mass matrix and $\psi$ is a column vector of spinors
of different neutrino masses\footnote{In this section, greek (latin) indices refer
to general curvilinear (locally inertial) coordinates.}. The vierbein fields $e^{\mu}_a$
connect the general curvilinear
and locally inertial sets of coordinates. The spinorial connection $\Gamma_{\mu}$ is
accordingly defined by
\begin{equation}
\label{eqn:spinconn}
\Gamma_{\mu}\,=\,\frac{1}{8}\left[\gamma^b,\gamma^c\right]\hspace{-0.5mm}e^{\nu}_{b}\nabla_\mu e_{c\hspace{0.2mm}\nu}\,.
\end{equation}
Using the relation
\begin{equation}
\label{tot}
\gamma^a\hspace{-0.9mm}\left[\gamma^b, \gamma^c\right]\,=\,2\eta^{ab}\gamma^c\,-\,2\eta^{ac}\gamma^b\,-\,
2\hspace{0.1mm}i\hspace{0.1mm}\epsilon^{abcd}\gamma^5\gamma_d\,,
\end{equation}
we find that the only non-vanishing contribution of the spin connection is
\begin{equation}
\gamma^a\hspace{0.1mm}e^{\mu}_a\Gamma_\mu\,=\,\gamma^a\hspace{0.1mm}e^{\mu}_a\left\{i\hspace{0.1mm}A_{G\mu}\left[-\,{\left(-g\right)}^{-1/2}\frac{\gamma^5}{2}\right]\right\}\,,
\end{equation}
where  we have denoted by $\epsilon^{abcd}$ the totally antisymmetric tensor with $\epsilon^{0123}\,=\,+1$ and
\begin{equation}
\label{eqn:A}
A_{G}^{\mu}\,=\,\frac{1}{4}{\left(-g\right)}^{1/2}e^{\mu}_a\epsilon^{abcd}\left(e_{b\sigma,\nu}\,-\,e_{b\nu,\sigma}\right)e^{\nu}_c e^{\sigma}_d\,,
\end{equation}
with $g\,\equiv\,\mathrm{det}\,g_{\mu\nu}$. 
%
Following Ref.~\cite{Cardall}, the generalized flavor neutrino state can now be written as
\begin{equation}
\label{eqn:genstate}
|\psi_\alpha\rangle\,=\,\sum_{k=1,2} U_{{\alpha k}}\hspace{0.1mm}(\theta)\hspace{0.1mm}e^{i\Phi}|\nu_k\rangle\,,\quad \alpha\,=\,e, \mu\,,
\end{equation}
where $U_{{\alpha k}}\hspace{0.1mm}(\theta)$ is the generic element of the Pontecorvo matrix in Eq.~(\ref{PMM}). The
neutrino oscillation phase is given by
\begin{equation}
\label{eqn:phaseneu}
\Phi\,=\,\int_{A}^{B} P_\mu\hspace{0.3mm} \frac{dx^\mu_{null}}{d\lambda}\hspace{0.3mm}d\lambda\,,
\end{equation}
where $P_\mu$ is the four-momentum operator that generates the spacetime
translation of the mass eigenstates and $\frac{dx^\mu_{null}}{d\lambda}$ is the null tangent vector to the
neutrino worldline $x^{\mu}$, parameterized by $\lambda$. For diagonal
metrics, denoting with $dl$ the differential proper distance at constant time,
we have
\begin{equation}
\label{eqn:lambd}
d\lambda\,=\,dl{\left[g_{00}\hspace{0.2mm}{\left(\frac{dx^0}{d\lambda}\right)}^2\right]}^{-1/2}.
\end{equation}
The momentum operator $P_\mu$ can be derived from the
generalized mass-shell condition
\begin{equation}
\label{eqn:genmascond}
\left(P^{\mu}\,-\,A^\mu_{G}\gamma^5\right)\left(P_{\mu}\,-\,A_{G\mu}\gamma^5\right)\,=\,M^2.
\end{equation}
As in Sec.~III, by requiring neutrino mass eigenstates to be energy eigenstates with a common energy $E_0$ and
assuming the spatial components of $P^\mu$ and $\frac{dx^\mu_{null}}{d\lambda}$ to be antiparallel, in the relativistic approximation one has~\cite{Cardall}
\begin{equation}
\label{eqn:pmudxmu}
P_\mu\hspace{0.3mm}\frac{dx^\mu_{null}}{d\lambda}\,=\,-\left(\frac{M^2}{2}\,+\,\frac{dx^\mu_{null}}{d\lambda}A_{G\mu}\gamma^5\,\right),
\end{equation}
where we have neglected terms of  $\mathcal{O}(A_G^2)$ and $\mathcal{O}{(A_GM^2)}$.

Let us now apply Eqs.~(\ref{eqn:A}) and (\ref{eqn:phaseneu})
to the particular case of a uniformly accelerated frame. In the same fashion as the previous heuristic analysis, we restrict to
$1+1$ dimensions; with reference to the metric tensor Eq.~(\ref{eqn:linelementacframe}), the only non-trivial component
of the vierbein fields is thus given by
\begin{equation}
\label{eqn:vierb}
e^{\mu}_a\,=\,{f(a,x)}^{-1/2},\quad \mathrm{for}\hspace{1.2mm} \mu=a=0\,,
\end{equation}
where ${f(a,x)}$ is defined as in Eq.~(\ref{eqn:funct}).
Inserting Eq.~(\ref{eqn:vierb}) into Eq.~(\ref{eqn:A}), one directly
obtains $A^{\mu}_G\,=\,0$, yielding
\begin{equation}
\label{P}
P_\mu\hspace{0.3mm}\frac{dx^\mu_{null}}{d\lambda}\,=\,-\frac{M^2}{2}\,.
\end{equation}
By use of  Eqs.~(\ref{eqn:lambd}) and (\ref{P}),
the phase in Eq.~(\ref{eqn:phaseneu}) then becomes
\begin{equation}
\label{tildphi}
\widetilde\Phi\,=\,\int_{A}^{B} P_\mu\hspace{0.3mm} \frac{dx^\mu_{null}}{d\lambda}\hspace{0.2mm}d\lambda\,=\,-\int_{A}^{B} \frac{M^2}{2\hspace{0.3mm}E_0}\hspace{0.2mm}{f(a, x)}^{1/2} dx\,,
\end{equation}
where we have exploited Eq.~(\ref{eqn:localenergy})
and the definition of proper distance introduced above.
After the mass operator in Eq.~(\ref{eqn:genstate}) has acted on $|\nu_k\rangle$,
we obtain
\begin{equation}
\widetilde\Phi_k\,=\,-\int_{A}^{B}\frac{m^2_k}{2\hspace{0.3mm}E_0}\hspace{0.2mm}{f(a, x)}^{1/2}dx\,, \quad k\,=\,1,2\,,
\end{equation}
that is exactly the same expression derived in Eq.~(\ref{eqn:covphaseapprox}).

\section{Applications}
In this section, we analyze some illustrative physical applications of our result. We begin by discussing the phenomenological implications of Eq.~(\ref{eqn:finalform}) in the framework of atmospheric neutrinos. In this case, mimicking the metric of a stationary observer on Earth with the one in Eq.~(\ref{eqn:linelementacframe}) and exploiting the equivalence principle, we can estimate the corrections induced by gravity to the probability of neutrino oscillations (we stress again that we are not concerned with effects of Earth's rotation). Then, we present a gedanken experiment in which these corrections are evaluated in more exotic regimes. We remark that, in both cases,
the condition $a\hspace{0.2mm}x\ll 1$ is satisfied.

\subsection{Earth's gravity effects on atmospheric neutrinos}
In the context of the atmospheric neutrino problem, it is known that flavor oscillations
can be faithfully analyzed using a simplified two-generations model, since they largely occur between muonic
and tauonic flavors ($\nu_\mu\leftrightarrow\nu_\tau$)\footnote{This happens because the mixing angle $\theta_{13}$
is much smaller than the others, and
two of the neutrino
mass states are very close in mass compared to the third ($\Delta m^2_{21}\ll\Delta m^2_{32}\approx\Delta m^2_{31}$ in the normal mass hierarchy)~\cite{Tanabashi}.}. 

Atmospheric neutrinos are produced in hadronic showers resulting from the interaction of cosmic rays with nuclei in the atmosphere.
Typical flight distances in experiments involving these neutrinos range from $10^2$ km (for neutrinos \emph{downward-going} from an interaction
above the detector) to more than $10^4$ km
(for neutrinos \emph{upward-going} from collisions on the other side of the Earth).
We restrict to the first case, where no background matter effect occurs.

Consider a detector comoving with the Earth: by restoring proper units in Eqs.~(\ref{pontec}) and (\ref{eqn:finalform}), a straightforward calculation then leads to
\begin{equation}
\label{Pvalue}
\big|P_{\nu_\mu\rightarrow\nu_\tau}\,-\,\widetilde P_{\nu_\mu\rightarrow\nu_\tau}\big|\,=\,\mathcal{O} \hspace{-0.3mm}\left(10^{-15}\right),
\end{equation}
where we have indicated with $P$ ($\widetilde P$) the oscillation probability as measured by the inertial (accelerated) observer. To numerically evaluate Eq.~(\ref{Pvalue}), we have set a neutrino mean flight path $L_p\sim10^2\,{\mathrm{km}}$, an acceleration of the order of Earth's gravity, $a\sim10\,{\mathrm{m/s}}^2$,  $\Delta m_{\mathrm{atm}}^2\sim 10^{-3}\,{\mathrm{eV}}^2$, $E_0\sim 1\,\mathrm{GeV}$ and maximal mixing $\theta_{atm}\approx \pi/4$~\cite{Tanabashi}.

 The obtained correction is far below the uncertainty on the current best-fit value of the oscillation probability $P_{\nu_\mu\rightarrow\nu_\tau}$, thus preventing any possibility of detecting gravitational effects on atmospheric neutrino oscillations at present. Future experiments, however, may give new insights in this direction.

\subsection{Neutrino oscillations in extreme acceleration regimes: a gedanken experiment}
We now propose a gedanken experiment in order to test our formalism in astrophysical regimes. In this framework, it is  reasonable to expect a larger contribution of gravitational effects on the oscillation probability, due to the extremely high accelerations that might be reached in this case.

As proof of this, let us consider an ideal accelerated detector in proximity of a high-density object; by way of illustration, we focus on the case of Sirius B,
the nearest (known) white dwarf to the Earth.
It is known that the gravity on the surface of this star is of the order of $10^6\, \mathrm{m/s^2}$~\cite{wd}. For such an acceleration, using Eqs.~(\ref{pontec}) and (\ref{eqn:finalform}), we obtain
\begin{equation}
\label{resu}
\big|P_{\nu_\alpha\rightarrow\nu_\beta}\,-\,\widetilde P_{\nu_\alpha\rightarrow\nu_\beta}\big|\,=\,\mathcal{O}\hspace{-0.3mm} \left(10^{-4}\right),
\end{equation}
where we have set $L_p\sim10^4\,{\mathrm{km}}$ and, as for the previous case, $\Delta m^2\sim10^{-3}\,{\mathrm{eV}}^2$, $E_0\sim 1\,\mathrm{GeV}$ and maximal mixing $\theta\approx \pi/4$. It is worth observing that the chosen value for the oscillation length still allows us to work with the metric of Eq.~(\ref{eqn:linelementacframe}) for the given acceleration.

As predicted, inertial effects may not be completely negligible in this case. However, it is worth saying that experiments like the one above considered are far from being viable nowadays. Indeed, it would be technically cumbersome to build a detector capable of withstanding the mechanical stress arising in those  regimes without breaking; on the other hand, even if it were possible, then the problem would arise of how to send and retrieve a probe from the surface of such remote sources (Sirius B, for example, lies at a distance of $8.60$ light-years away from the Sun).

Notwithstanding these technical difficulties, some of the implications of the result Eq.~\eqref{resu} in the physics of neutrino oscillations will be discussed in the next section.

\section{Discussion and conclusion}
We have analyzed neutrino flavor oscillations from the point of view of a uniformly
accelerated observer. Corrections to the standard result
have been derived by use of Stodolsky covariant definition
of neutrino quantum phase.
Relying on phenomenological considerations,
we have restricted our discussion to relativistic neutrinos,
so that a plane-wave treatment could be applied.
In order to realize how acceleration affects the usual Pontecorvo formula,
the formalism of neutrino
oscillations in curved spacetime has been used.
Within such a framework, we have found that inertial effects
are intimately related to the redshift of neutrino energy,
according to Ref.~\cite{Cardall}. Furthermore, it has
been pointed out that a separate ``acceleration phase" can be
extracted from the standard result only for small accelerations.

As a possible application of our analysis, we have
calculated the correction induced by Earth's gravity on the oscillation
probability of atmospheric neutrinos. In that case, simulating the metric of an observer comoving with the Earth with the one in Eq.~(\ref{eqn:linelementacframe}),  we have found that the contribution to the neutrino phase-shift is negligible,
thus leading to effects which are currently unmeasurable.
It is clear that the origin of this outcome can be traced back to the difficulty of detecting gravitational effects on oscillations in the weak-field regime, as it is near to the Earth.
On the other hand, in astrophysical regimes (e.g. outside a black hole or in proximity of pulsars)
we expect these corrections to be far more relevant (as also suggested by the analysis carried out for a white dwarf), resulting in a possible modification of the oscillation probability induced by gravity (see also Ref.~\cite{Miller:2013wta} for a quasi-classical treatment of neutrino oscillations in the gravitational field of a heavy astrophysical object). If confirmed, such an effect could be exploited for investigating the gravity-induced interactions that neutrinos may have experienced during their travel throughout the Universe, and thus the mass distribution of the Universe itself. Oscillations of neutrinos from supernovae and active galactic nuclei may be  valuable to search traces of space-time quantum foam~\cite{KlapdorKleingrothaus:2000fr}. Non-trivial implications may also arise in the context of supernova nucleosynthesis, and, in particular, in the production of heavy elements in neutrino-driven winds from proto-neutron stars and neutrino-induced nucleosynthesis in outer shells of supernovae~\cite{Wu}. A further interesting scenario to explore is the r\^ole of neutrino oscillations in the generation of the rotational pulsar velocity in the presence of intense magnetic fields~\cite{Kusenko}. These aspects, however, will be investigated in future publications.

Aside from phenomenological aspects, we emphasize
that investigating inertial effects on neutrino mixing and oscillations may be
useful for clarifying a variety of controversial problems at a  theoretical level.
Recently, for instance, some concerns have been raised regarding the agreement between the decay rates of accelerated protons in the inertial and comoving frames when neutrino mixing is taken into account~\cite{Ahluwalia:2016wmf,Blasone:2018czm,Cozzella:2018qew}. 
Besides, the evolution of neutrinos in a background matter moving with a
linear acceleration has been analyzed in Ref.~\cite{Dvornikov}.
Relevant processes in non-inertial frames are studied also in condensed matter physics. In Ref.~\cite{Basu:2013kt}, in particular, an enhancement of the spin current for a linearly accelerating semiconductor system has been predicted.

Finally, we remark that our whole analysis has been performed
in the context of Quantum Mechanics.
Along the line of Refs.~\cite{Blasone:2017nbf,Blasone:2018byx},
it naturally arises the question of
how  the oscillation probability for an accelerated observer would appear
within the framework of Quantum Field Theory. Work is already in progress along this line~\cite{prepa}.

\end{document}